\documentclass[aps,prd,reprint,nofootinbib,floatfix]{revtex4-1}
\usepackage{times}		
\usepackage{booktabs}
\usepackage{placeins}
\pdfoutput=1
\usepackage{amssymb,amsmath,graphics,epsfig}
\usepackage{amsmath}

\def\apjl{{{\apj} Lett.}}
\def\mnras{{Mon. Not. R. Astron. Soc.}}
\def\apss{{Astrophysics and Space Science}}
\def\aap{{Astronomy and Astrophysics}}

\newcommand{\rem}[1]{ }
\newcommand{\beq}{\begin{equation}}
\newcommand{\eeq}{\end{equation}}
\newcommand{\bea}{\begin{eqnarray}}
\newcommand{\eea}{\end{eqnarray}}

\begin{document}
\title{Quasi-nonlinear approach to the Weibel instability}


\author{Mikhail V. Medvedev} 
\altaffiliation{On sabbatical leave from the Department of Physics and Astronomy, University of Kansas, Lawrence, KS 66045}
\affiliation{Institute for Theory and Computation, Harvard University, Cambridge, MA 02138}

\begin{abstract}
Astrophysical and high-energy-density laboratory plasmas often have large-amplitude, sub-Larmor-scale electromagnetic fluctuations excited by various kinetic-streaming or anisotropy-driven instabilities. The Weibel (or the filamentation) instability is particularly important because it can rapidly generate strong magnetic fields, even in the absence of seed fields. Particles propagating in collisionless plasmas with such small-scale magnetic fields undergo stochastic deflections similar to Coulomb collisions, with the magnetic pitch-angle diffusion coefficient representing the effective ``collision'' frequency. We show that this effect of the plasma ``quasi-collisionality'' can strongly affect the growth rate and evolution of the Weibel instability in the deeply nonlinear regime. This result is especially important for understanding cosmic-ray-driven turbulence in an upstream region of a collisionless shock of a gamma-ray burst or a supernova. We demonstrate that the quasi-collisions caused by the fields generated in the upstream suppress the instability slightly but can never shut it down completely. This confirms the assumptions made in the self-similar model of the collisionless foreshock. 
\end{abstract}

\maketitle

\section{Introduction}

High-amplitude small-scale electromagnetic turbulence is ubiquitous in high-energy density plasmas. It is often excited at kinetic scales by the anisotropy of the particle distribution, including their counter-streaming motions. Particularly interesting are the Weibel-type streaming instabilities (often called the `filamentation instabilities') \citep{weibel59, fried59}, which is capable of producing strong magnetic fields. Such magnetic ``Weibel turbulence" is a common feature of astrophysical and space plasmas, e.g., collisionless shocks in gamma-ray bursts (GRBs) and supernova remnants, and other weakly magnetized plasmas \citep{medvedev+99, nishikawa+03, silva+03, frederiksen+04, nishikawa+05, spitkovsky+08, nishikawa+09, keshet+09, medvedev+z09, kamble+14}, sites of magnetic reconnection \citep{swisdak+08, liu+09} and others. Weibel fields play a critical role in high-intensity laser plasmas studies in laboratory facilities such as the National Ignition Facility, OmegaEP, Hercules, Trident, and others \citep{tatarakis+03, ren+04, huntington+12, mondal+12, kugland+12, huntington+15, park+15}. Experimental and numerical studies of the instability itself and of non-relativistic collisionless shocks, motivated by the proposed idea of ``a GRB in a lab"  \citep{medvedev07, medvedev08, medvedev+s09}, demonstrate crucial importance of small-scale fields \citep{medvedev+05, bret+05, bret+08, achterberg+07a, achterberg+07b, lemoine+09, medvedev+s09, medvedev11, fiuza+12, kugland+12, huntington+15, park+15}. These fields are also important for the fusion energy sciences and the inertial confinement concept \citep{ren+04, tatarakis+03}.

Despite much variation in the plasma conditions in which the Weibel-like or filamentation instabilities are excited, most of these plasmas have one thing in common. While being collisionless or weakly collisional, that is the binary Coulomb collisions are rare, these environments display phenomena that resemble conventional collisional interactions. For instance, relativistic  collisionless shocks in GRBs have the particle mean-free-paths being only tens of plasma skin depths. This happens because the small-scale fields in the Weibel turbulence vary on scales smaller than or comparable to the characteristic curvature scale of the particles traversing the field, i.e., the particle Larmor radius. The particle trajectory through these turbulent fields will, consequently, never form a well-defined Larmor circle. Furthermore, even if the average magnetization is substantial and the particle mostly moves along its Larmor orbit, the large fluctuations of the field strength can break the particle's adiabatic invariant in certain locations, thus leading to pitch-angle scattering. Hereafter, we colloquially refer to these phenomena as ``quasi-collisional'' \citep{keenan+13,keenan+15,keenan+16}.

\section{Quasi-collisionality of sub-Larmor turbulence}

Consider a particle moving with velocity ${\bf v}$ through a random, mean-free $\langle {\bf B} \rangle = 0$, small-scale magnetic field which is statistically homogeneous and isotropic. The Lorentz force, ${\bf F}_L=(e/c)\,{\bf v\times B}$, acting on the particle is random, hence particle's velocity and acceleration vectors vary stochastically, leading to a random (diffusive) trajectory. We define the field turbulence to be ``small-scale'' when the effective particle's Larmor radius, $r_L$, is greater than, or comparable to, the characteristic correlation scale of the magnetic field, $\lambda_B$, i.e., $r_L\gtrsim \lambda_B$, where $r_L =  (\Gamma^2-1)^{1/2} m c/e \langle B_\perp^2 \rangle^{1/2}$ with $\langle B_\perp^2 \rangle^{1/2}$ being the rms component of the magnetic field perpendicular to the particle's velocity vector, $m$ being the mass, $c$ being the speed of light, $e$ being the electric charge, and $\Gamma$ being the particle's Lorentz factor.

The deflection angle of the velocity, assuming small deflections, is approximately equal to the ratio of the change in the particle's transverse momentum to its initial transverse momentum. As the particle passes though a coherent patch of magnetic field of length $\lambda_B$ it experiences the force $F_L\sim e B v_\perp/c$ over the transit time $\tau_\lambda\sim\lambda_B/v_\perp$. The change in transverse momentum is, therefore ${\Delta}p_\perp \sim F_L \tau_\lambda \sim e(B/c)\lambda_B$, so that the deflection angle becomes $\alpha_\lambda \approx {\Delta}p_\perp/p_\perp \sim e(B/c)\lambda_B/(\Gamma m_e v_{\perp})$, where the particle's total transverse momentum is $p_\perp \sim \Gamma m v_\perp$. The subsequent deflection in the next patch is uncorrelated and, thus, will be in a random direction. Hence the particle motion is diffusive. 

The pitch-angle diffusion coefficient is defined as the ratio of the square of the deflection angle in a coherent patch to the transit time over this patch, that is
\begin{equation}
D_{\alpha\alpha} \sim \frac{\alpha_\lambda^2}{\tau_\lambda}\sim \frac{e^2 \langle B^2 \rangle \lambda_B}{\Gamma^2 m^2 c^2 \langle v_{\perp}^2 \rangle^{1/2} },
\label{Daa}
\end{equation}
where a volume-averaged square magnetic field, $\langle B^2 \rangle$, and perpendicular rms velocity, $\langle v_{\perp}^2 \rangle^{1/2}$, have been substituted for $B^2$ and $v_{\perp}$. 

Although the correlation length, $\lambda_B$, lacks a formal definition, it is often defined via the two-point autocorrelation tensor \citep{biswas+02}. The correlation length tensor, which is formally dependent on path and time, is defined as
\begin{equation}
\lambda^{ij}_B(\hat{\bf r}, t) \equiv \int_{0}^\infty \! \frac{R^{ij}({\bf r}, t)}{R^{ij}(0, 0)}  \, \mathrm{d}r,
\label{corr_l}
\end{equation}
where
\begin{equation}
R^{ij}({\bf r}, t) \equiv \langle {B}^i({\bf x}, \tau){B}^j({\bf x} + {\bf r}, \tau + t) \rangle_{{\bf x}, \tau}
\label{corr_tensor}
\end{equation}
and we make no distinction between co-variant and contra-variant components. Let ${\bf B}_{{\bf k}, \Omega}$ be the spatial and temporal Fourier transform of the magnetic field, ${\bf B}_{{\bf k},\Omega} = \int \! {\bf B}({\bf x}, t)e^{-i({\bf k}\cdot{\bf x} - \Omega{t})} \, \mathrm{d} {\bf x} \mathrm{d}t$,
where ${\bf k}$ and $\Omega$ are the corresponding wave vector and frequency, respectively. Then it is often convenient to define a complementary spectral correlation tensor for the field $\Phi_{ij}({\bf k}, \Omega)$, such that
\begin{equation}
R_{ij}({\bf r}, t) = (2\pi)^{-4}\int \Phi_{ij}({\bf k}, \Omega) e^{i{\bf k}\cdot{\bf r} -i\Omega{t}} \, \mathrm{d}{\bf k}\, \mathrm{d}{\Omega}.
\label{corr_tensor_def_Phi}
\end{equation}
The spectral correlation tensor, $\Phi_{ij}({\bf k}, \Omega)$, naturally connects statistical properties of the field to its spectral characteristics.

In order to proceed further, one needs to know the full three-dimensional spectrum of the magnetic field generated by the Weibel instability. In general, Weibel turbulence is anisotropic. One can expect however that in the deeply nonlinear regime, it may tend to isotropy. Thus, we are making now a strong simplifying assumption of the isotropy and time-independence. Together with ${\bf \nabla}\cdot{\bf B} = 0$, these assumptions require the spectral correlation tensor to be of the form 
\begin{equation}
\Phi_{ij}({\bf k}, \Omega) = \frac{1}{2V}\left|{\bf B}_k\right|^2\left(\delta_{ij} - \hat{k}_i\hat{k}_j\right)2\pi\delta(\Omega),
\label{spec_tensor}
\end{equation}
where $V$ is the volume of the space considered, $\hat{\bf k}$ is the unit vector in the direction of the wave vector, and $\delta_{ij}$ is the Kronecker delta. The normalization has been chosen such that $\sum{R}_{ii}(0, 0) = \langle B^2 \rangle_{{\bf x}, \tau} = \langle B^2 \rangle$. 

Since only the component of the magnetic field perpendicular to the particle trajectory is relevant, we choose an integration path in Eq. (\ref{corr_l}) to be along ${\bf v_\perp}$ and only consider a transverse magnetic field component, and choose ${\bf r} = x\hat{\bf x}$ and $i=j=z$. The magnetic field correlation length becomes
\begin{equation}
\lambda_B \equiv \lambda^{zz}_B(\hat{\bf x}, t)  = \int_{0}^\infty \! \frac{R^{zz}(x\hat{\bf x}, t)}{R^{zz}(0, 0)}  \, \mathrm{d}x.
\label{corr_l_def}
\end{equation}
Using Eqs. (\ref{corr_tensor})--(\ref{corr_l_def}), noting that ${\bf B}_{\bf k}$ is only a function of $|{\bf k}| \equiv k$ and integrating over $\mathrm{d}x$ and all solid angles in $\mathrm{d}{\bf k}$, we finally obtain \citep{keenan+15}
\begin{equation}
\lambda_B = \frac{3\pi}{8}\frac{\int_{0}^\infty \! k{|{\bf B}_k|^2}\, \mathrm{d}k}{\int_{0}^\infty \! k^2{|{\bf B}_k|^2}\, \mathrm{d}k}.
\label{corr_l_div}
\end{equation}
By its physical meaning, the correlation length represents a characteristic wave number of turbulence, $\lambda_B \approx k_B^{-1}$. 

Finally, the  collision frequency is defined as the inverse time during which the rms pitch-angle deflection becomes of order one radian, thus $D_{\alpha\alpha}\nu_\textrm{eff}^{-1} =\langle\alpha^2\rangle \sim 1$. Using Eqs. (\ref{Daa}) and (\ref{corr_l_div}), we have
\begin{equation}
\nu_\textrm{eff}=D_{\alpha\alpha} = \frac{3\pi}{8}\sqrt{\frac{3}{2}}\frac{e^2}{m^2 c^2}
\left(\frac{\int_{0}^\infty \! k{|{\bf B}_k|^2}\, \mathrm{d}k}{\int_{0}^\infty \! k^2{|{\bf B}_k|^2}\, \mathrm{d}k}\right)
\frac{\langle B^2 \rangle}{\Gamma^2 v_{th}}, 
\label{nueff}
\end{equation}
where we used that $\langle v_{\perp}^2 \rangle = (2/3)v_{th}^2$ for an isotropic particle distribution with $v_{th}$ being the characteristic thermal speed.

\section{Quasi-collisional Weibel instability}

Let us consider the electron Weibel instability in the uniform, static, charge-neutralizing background of protons. (Numerical simulations indicate that the ion-driven Weibel instability proceeds similar to the electron-driven one, even though the electrons need not be forming a uniform charge-neutralizing background.) For simplicity, we assume the distribution of the electrons is represented by two cold, counter-propagating streams. Here we follow the derivations in Refs. \citep{honda04,califano+97}. The governing equations are 

\bea
& \displaystyle \frac{\partial n_a}{\partial t}-\nabla\cdot{\bf j}_a=0, & \label{eq1}\\
& \displaystyle \frac{\partial {\bf p}_a}{\partial t}+{\bf v}_a\cdot\nabla{\bf p}_a = 
-({\bf E}+{\bf v}_a\times{\bf B})-\nu_\textrm{eff}({\bf p}_a-{\bf p}_{\bar a}), & \label{eq2}\\
& \displaystyle \nabla\times{\bf E}=-\frac{\partial {\bf B}}{\partial t}, & \label{eq3}\\
&  \displaystyle \nabla\times{\bf B}=\frac{\partial {\bf E}}{\partial t} + \sum_a\ {\bf j}_a, & \label{eq4}\\
&  \displaystyle \nabla\cdot{\bf E}=1- \sum_a\ n_a, & \label{eq5}
\eea
where ${\bf j}_a=-n_a{\bf v}_a$ and ${\bf v}_a={\bf p}_a/\sqrt{1+p_a^2}$. The index $a=1,2$ denotes the the two counter-streaming electron populations and $\bar a$ denotes the counterpart of $a$, that is $\bar a=2$ if $a=1$ and $\bar a=1$ if $a=2$. Hereafter, the densities are normalized by the uniform density $n_0$, velocities by the speed of light $c$ and frequencies by the plasma frequency $\omega_{p}=(4\pi e^2 n_0/m)^{1/2}$. Obviously the first (continuity) equation in the system is derivable from the last two (Ampere and Poisson) equations. 

Now, we assume that the electron streams are initially propagating along $x$-direction, that is ${\bf v}_{0,a}=v_{0,a}\hat{\bf x}$. The Weibel instability is also characterized by current neutrality, $\sum_a n_{0,a}v_{0,a}=0$, thus there is no initial magnetic field. We assume that the growing magnetic field will be in the $z$-direction, ${\bf B}=B\hat{\bf z}$ and the perturbed velocities and electric fields lie in the orthogonal, $x$-$y$-plane.The Weibel instability is a transverse instability, so we dismiss longitudinal electrostatic perturbations by considering perturbations of the form $e^{(ik_y y-i\omega t)}$.

\begin{figure*}[t]
\vskip1em
\centering
\includegraphics[scale = 0.65]{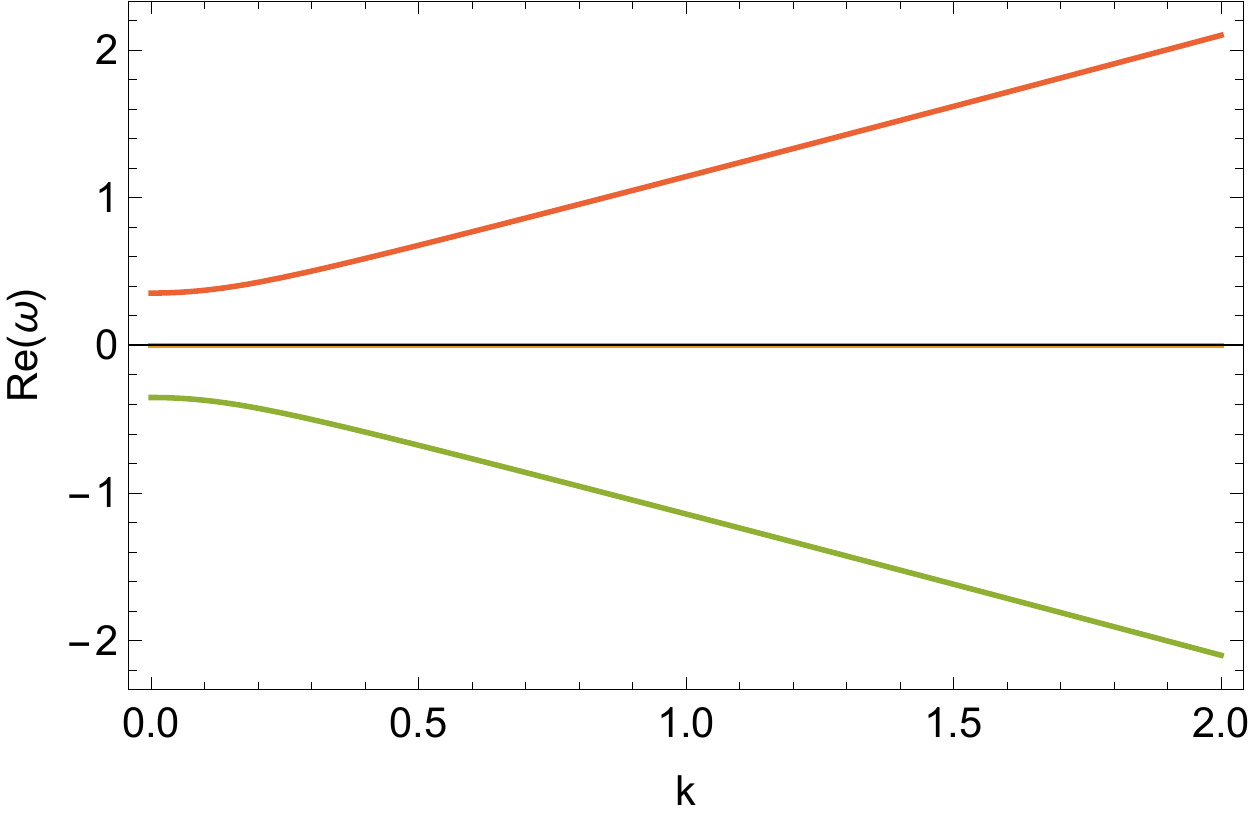}
\includegraphics[scale = 0.67]{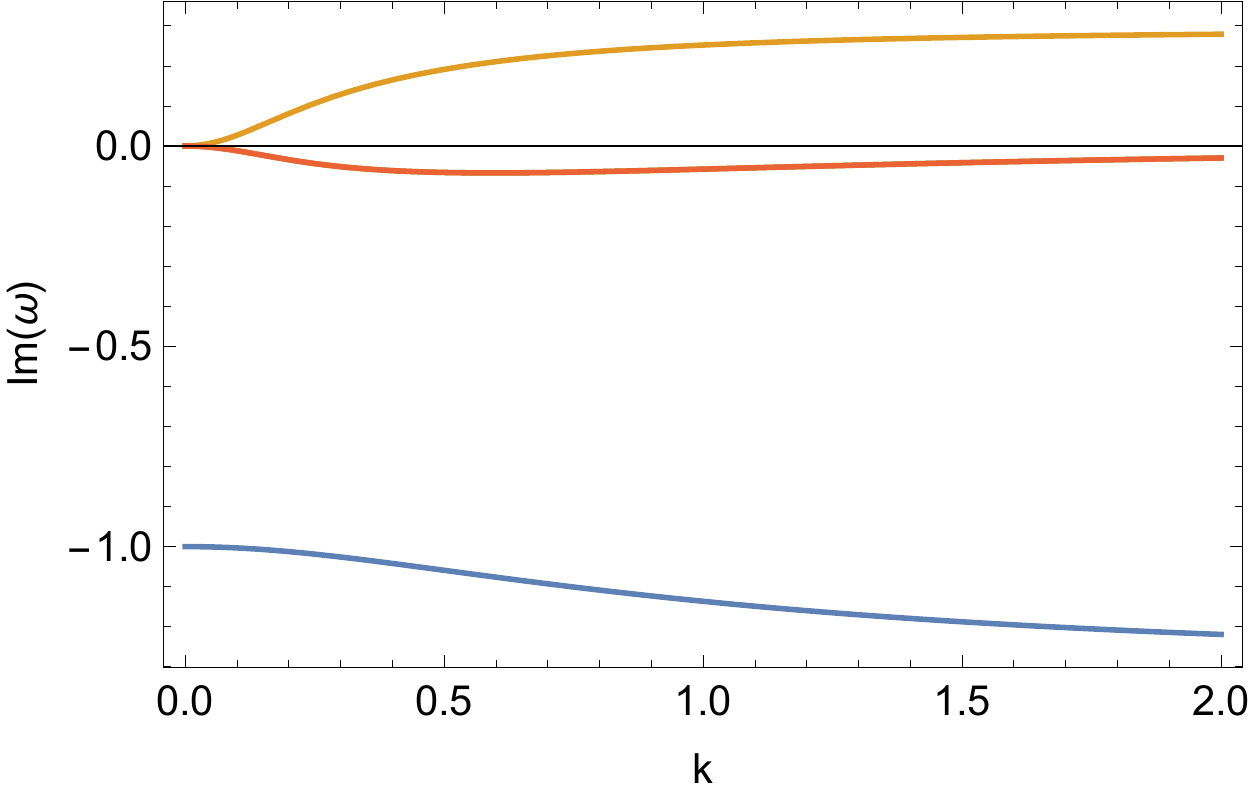}
\caption{Dispersion curves, $\Re(\omega)$ and $\Im(\omega)$ vs $k$, obtained by equating the term in the square brackets in Eq. (\ref{disp}) to zero. Two modes have large real frequencies and are slightly damped by $\nu_\textrm{eff}$. The other two modes have zero real frequencies; one is damped and one is unstable, which is the Weibel mode.}
\label{f:disp} 
\end{figure*}
\begin{figure}
\vskip1em
\centering
\includegraphics[scale = 0.65]{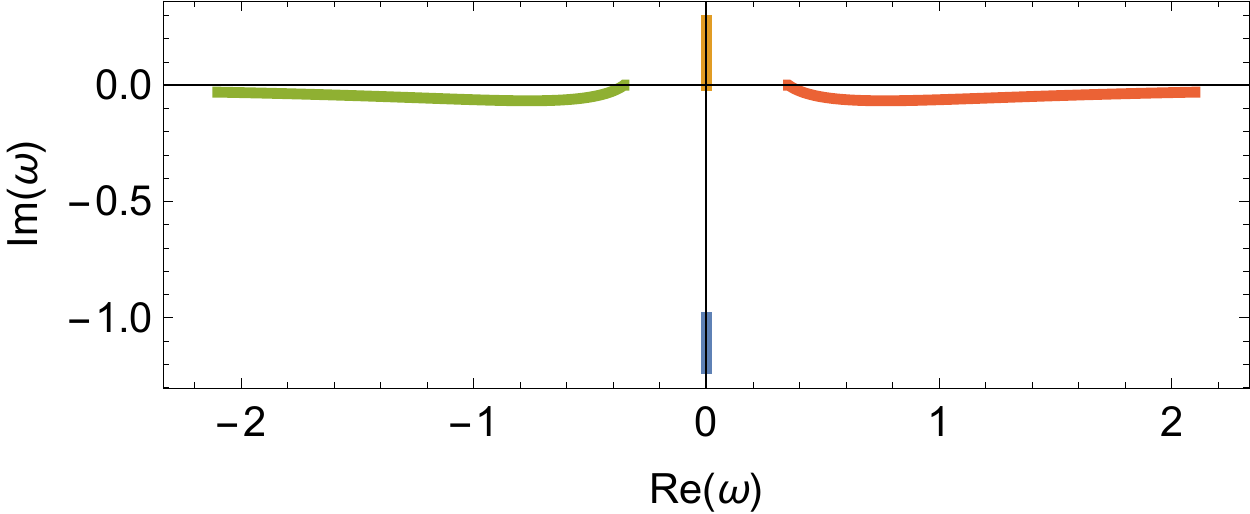}
\caption{Roots in the complex $\Re(\omega)$-$\Im(\omega)$ plane for $0\le k\le2$. The color code is the same as in Fig. \ref{f:disp}. }
\label{f:roots} 
\end{figure}

Upon solving the system of linearized equations (\ref{eq1})-(\ref{eq5}), one obtains the following dispersion relation (see Ref. \citep{honda04} for the general treatment):
\beq
\omega^2(1-A_1)(1-A_2)-k^2(1-A_1)(1+A_3)+A_4=0,
\label{disp1}
\eeq
where
\bea
A_1 &=& \sum_a\frac{n_{0,a}}{\Gamma_{0,a}\omega^2},\\
A_2 &=& \sum_a\frac{n_{0,a}}{\Gamma_{0,a}^3\omega^2},\\
A_3 &=& \sum_a\frac{n_{0,a}v_{0,a}^2}{\Gamma_{0,a}\omega'^2},\\
A_4 &=& \left(\sum_a\frac{n_{0,a}v_{0,a}}{\Gamma_{0,a}\omega^2}\right)
\left(\sum_a\frac{n_{0,a}v_{0,a}}{\Gamma_{0,a}\omega'^2}\right),\\
\omega'^2 &=& \omega^2\ \frac{\omega+2i\nu_\textrm{eff}}{\omega+i(1+v_{0,\bar a}/v_{0,a})\nu_\textrm{eff}}
\eea
and $\Gamma_{0,a}=\left(1-v_{0,a}^2\right)^{-1/2}$ is the Lorentz factor.

Analysis of this equation is still cumbersome, so we further simplify equations by assuming that the interpenetrating electron streams are of the same densities, $n_{0,1}=n_{0,2}=0.5$, and therefore the same speed, i.e., $v_{0,1}=-v_{0,2}$. We introduce $v_0=|v_{0,a}|$ so that $\Gamma_0=(1-v_0^2)^{-1/2}$. The dispersion relation (\ref{disp1}) reduces to
\beq
\left(\omega^2-\Gamma_0^{-1}\right)\left[\omega'^2\left(\omega^2-\Gamma_0^{-3}\right)
-k^2\left(\omega'^2+v_0^2\Gamma_0^{-1}\right)\right]=0, 
\label{disp}
\eeq
where 
\beq
\omega'^2=\omega(\omega+2i\nu_\textrm{eff}).
\eeq

The eigenmode that factored out represents the standard relativistic plasma oscillation, $\omega=\pm\Gamma_0^{-1/2}$, where $\omega$ has a vanishing imaginary part. Obviously, it is not affected by the effective collisionality.

The term in the square brackets of Eq. (\ref{disp}) yields four solutions, one of them corresponds to the Weibel instability. The roots and dispersion curves are shown in Figs. \ref{f:disp}, \ref{f:roots}. These results differ from those obtained in Ref. \citep{honda04}, where the parameter like $\nu_\textrm{eff}/\omega$ in our notations was treated as a real-valued constant, which is incorrect. Among the four roots, two modes have large real frequencies and are slightly damped by collisions. The other two modes have vanishing real frequencies; one mode is damped and one is purely growing. The latter, unstable mode is the Weibel instability. Note that in our treatment of the cold plasma, the mode is unstable for an arbitrarily large $k$. In reality, there is a maximum $k$, which depends on the thermal velocity spread, see Ref. \citep{achterberg+07a} for the extensive analysis and discussion.

Since, the Weibel instability is a purely growing mode in our analysis, we define the growth rate as $\gamma=i\omega$. It is, thus, a solution to the equation
\beq
\gamma(\gamma+2\nu_\textrm{eff})\left(\gamma^2+\Gamma_0^{-3}\right)
+k^2\left(\gamma(\gamma+2\nu_\textrm{eff})-v_0^2\Gamma_0^{-1}\right)=0.
\label{gamma}
\eeq

\begin{figure*}
\vskip1em
\centering
\includegraphics[scale = 0.65]{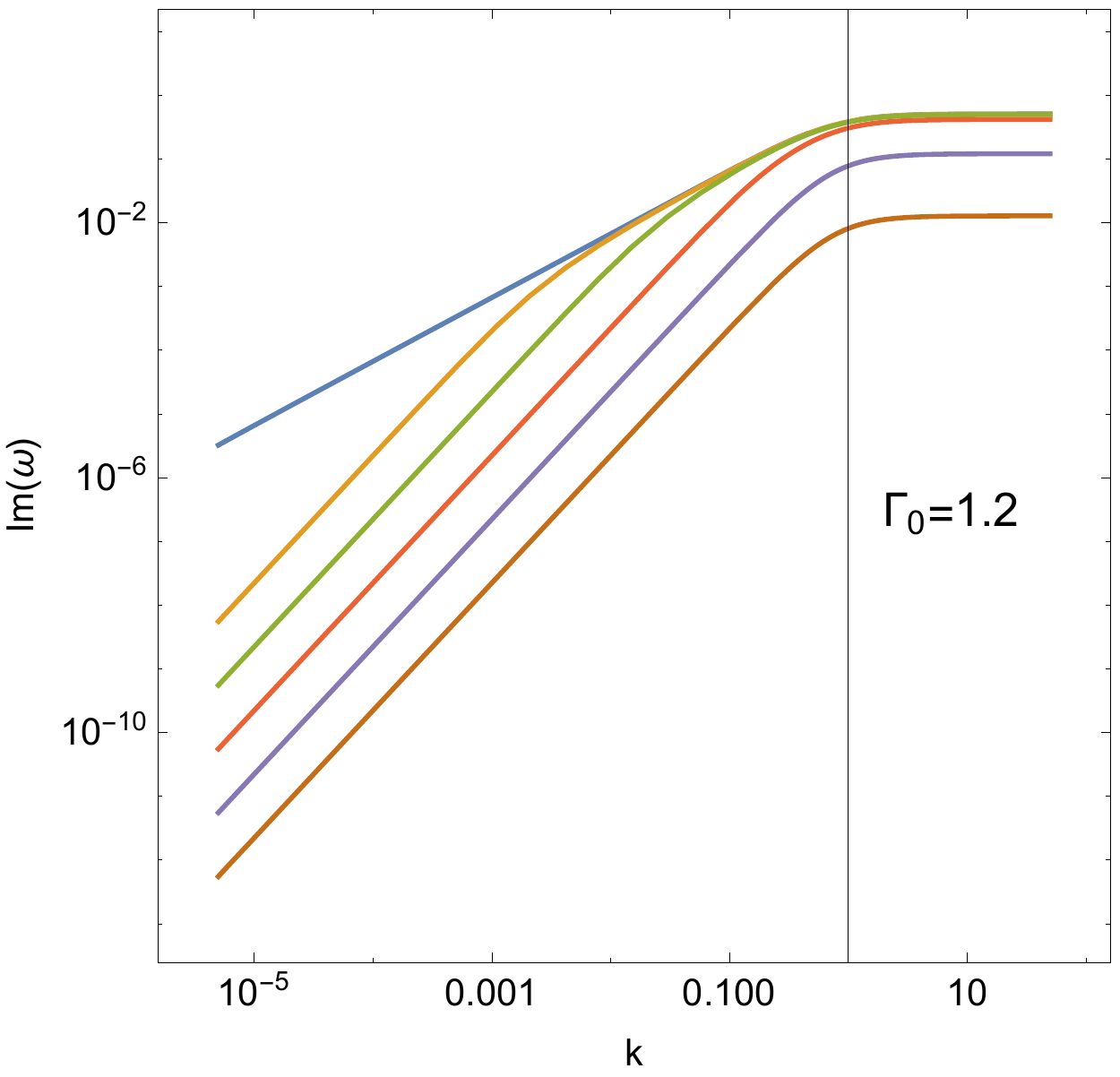}
\includegraphics[scale = 0.65]{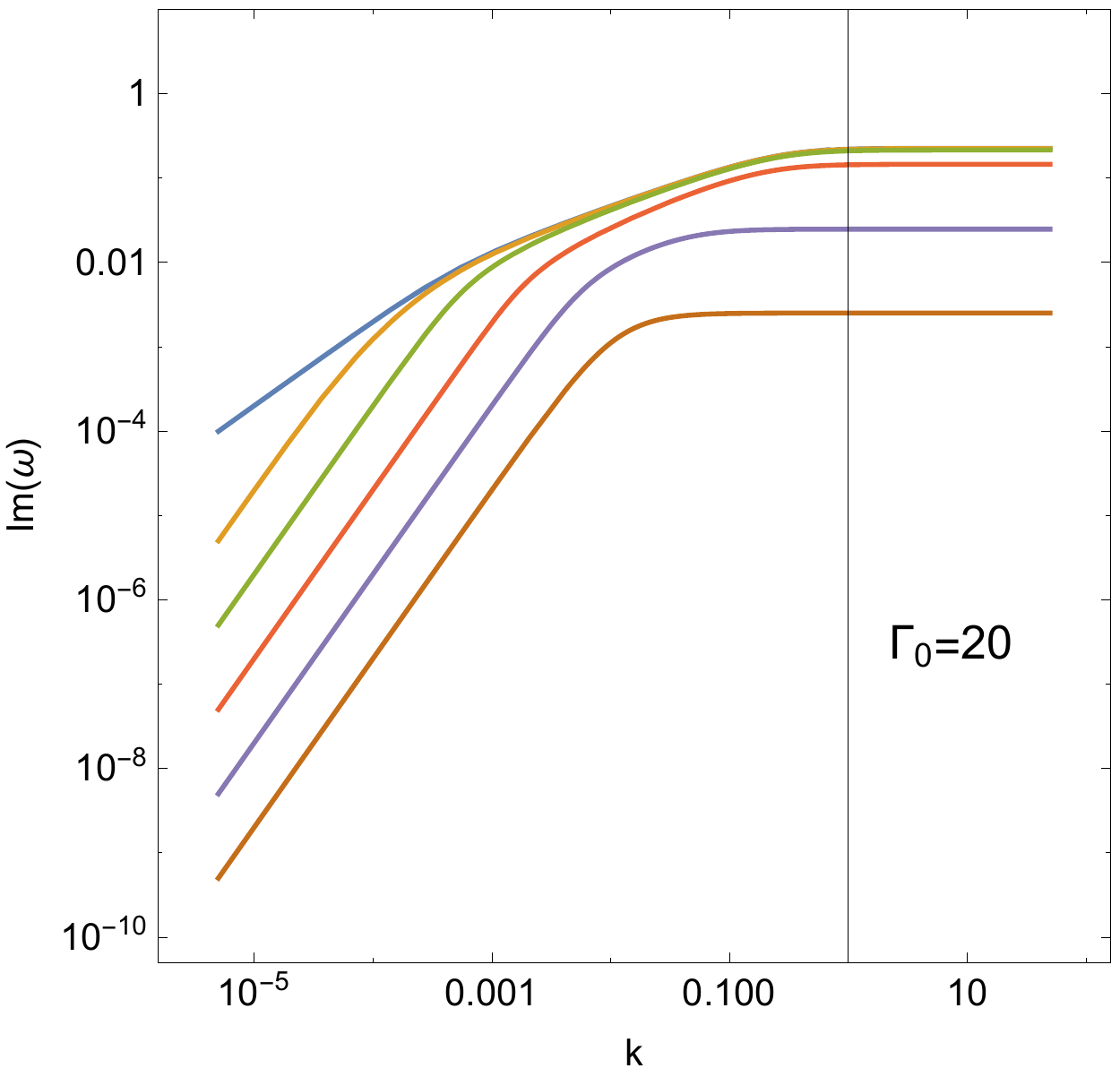}
\caption{Quasi-collisional Weibel instability growth rates in the non-relativistic and relativistic regimes, $\Gamma_0=1.2$ and  $\Gamma_0=20$, respectively. The curves, from top to bottom, correspond to $\nu_\textrm{eff}=0,\ 10^{-2}, 10^{-1}, 1, 10, 100$. The vertical line denotes where $k=1$.}
\label{f:gamma} 
\end{figure*}
The solutions to it are shown in Fig. \ref{f:gamma} for various values of the effective quasi-collisional frequency in the range $0.01\le \nu_\textrm{eff}\le100$. The classical Weibel dispersion relation  (with $\nu_\textrm{eff}=0$) is also shown for reference. 
It is seen that the growth rate is suppressed by collisionality but never goes to zero for any finite $\nu_\textrm{eff}$. For small quasi-collisionalities, $\nu_\textrm{eff}\ll1$, the small-$k$ regime is suppressed the most, where is becomes $\gamma\propto k^{2}$ instead of the classical $\gamma\propto k$ scaling, for $\Gamma_0\sim1$. In the relativistic limit, $\Gamma_0\gg1$, there appears a second break at low $k$, with the intermediate scaling $\gamma\propto k^{1/2}$ at $\Gamma_0^{-3/2}\ll k\ll \Gamma_0^{-1/2}$. These $\gamma(k)$ scalings can readily be obtained from Eq. (\ref{gamma}) by considering appropriate limits: $\gamma\ll \nu_\textrm{eff}$ and $\nu_\textrm{eff}\ll \gamma\ll \gamma_m$, together with $k\gg\Gamma_0^{-3/2}$ or $k\ll\Gamma_0^{-3/2}$ .The position of the main break, i.e., the minimum $k$ at which the growth rate is still close to the maximum is approximately $k_m\sim\Gamma_0^{-1/2}$. At large values of $\nu_\textrm{eff}$, the overall growth rate decreases as approximately $\gamma\propto1/\nu_\textrm{eff}$. 

The maximum growth rate can be obtained by observing that $\gamma\to\gamma_m\sim const.$ as $k\to\infty$. Eq. (\ref{gamma}) in this limit becomes 
\beq
\gamma(\gamma+2\nu_\textrm{eff})-v_0^2/\Gamma_0 = 0,
\eeq
which admits the positive solution
\bea
\gamma_m &=& \sqrt{\nu_\textrm{eff}^2+v_0^2/\Gamma_0}-\nu_\textrm{eff} \nonumber\\
&\approx& \left\{
\begin{array}{ll}
\gamma_0-\nu_\textrm{eff}, & \textrm{ if }\ \nu_\textrm{eff}\ll\gamma_0,\\
\gamma_0^2/(2\nu_\textrm{eff}), & \textrm{ if }\ \nu_\textrm{eff}\gg\gamma_0,
\end{array}
\right.
\label{gammam}
\eea
where $\gamma_0=v_0/\sqrt\Gamma_0$ is the maximum growth rate of the classical Weibel instability. Note that the obtained growth rate depends on both the amplitude of the magnetic fluctuations, $\langle B^2\rangle$ and the effective correlation length of the field, $\lambda_B$.

\section{Discussion}

Results obtained in previous sections allow us to estimate the back-reaction of the generated fields on the instability. We stress, that our treatment differs from the standard quasilinear theory, in which the response of the particle distribution function is computed as a perturbation and then substituted back into the general dispersion relation. Our approach also differs from the conventional non-linear approach which considers the evolution of current filaments when the instability has already been saturated \citep{medvedev+05, achterberg+07b}. In contrast to standard approaches, we have considered here a case when the anisotropy of the particle distribution function is maintained in the unstable (streaming) state. In this case, the generated fields are affecting the instability via pitch-angle scattering of the streaming particles -- a clear-cut of ``quasi-collisions''. 

The back-reaction via quasi-collisions is of great importance for astrophysical collisionless shocks, both relativistic and non-relativistic, in GRBs and supernova remnants. At such shocks, an almost steady-state, yet unstable, particle distribution is formed in the vicinity of the shock front in the upstream region because of particle reflection and injection at the shock. Thus, the instability in the near-upstream region is manifestly in the ``quasi-nonlinear'' regime. Furthermore, as the shock evolves, it populates the far-upstream region with suprathermal particles and cosmic rays which generate the magnetic field on longer temporal and spatial scales \citep{keshet+09}. An analytical self-similar model of such a foreshock has been developed, but it assumed that the growth rate of the instability is not substantially modified by the self-generated fields \citep{medvedev+z09}. We can now check this assumption. 

For estimates, it is reasonable to assume that the characteristic scale of the turbulence is set by the low-$k$ modes with the maximum growth rate, thus $k_B\sim\Gamma_0^{-1/2}$. Restoring dimensional factors, we have
\beq
\lambda_B\sim(c/\omega_p)  \Gamma_0^{1/2}. 
\eeq
Then, from Eq. (\ref{nueff}) or (\ref{Daa}), one has
\beq
\nu_\textrm{eff}\sim\frac{\omega_p}{\Gamma_0^{1/2}\beta}\,\frac{v_0}{c},
\eeq
where we defined the generalized plasma $\beta$ as 
\beq
\beta\equiv\frac{\Gamma_0 n_0 (m v_0^2/2)}{\langle B^2\rangle/8\pi}. 
\eeq
From Eq. (\ref{gammam}), the quasi-nonlinear, amplitude-dependent growth rate becomes
\beq
\gamma_m(\beta)\sim\frac{\omega_p}{\Gamma_0^{1/2}}\,\frac{v_0}{c} 
\times\left\{
\begin{array}{ll}
1-\beta^{-1}, & \textrm{ if  } \beta\gg1,\\
\beta, & \textrm{ if  } \beta\ll1.
\end{array}
\right.
\eeq
If the unstable particle distribution is not maintained, the free energy of the instability is the {\em initial} particle distribution anisotropy. Then, the field energy should not exceed the kinetic energy of the particle streams, hence $\beta>1$ if $v_0\ll c$ and $\beta>1/2$ if $\Gamma_0\gg1$, so that $\nu_\textrm{eff}\lesssim\gamma_0$. 
Numerical simulations of the instability itself, as well as of the collisionless shocks, show similar results that the magnetic energy density does not usually exceed about 10\% of the kinetic energy density, i.e., $\beta\gtrsim10$. Thus, the influence of quasi-collisions induced by the self-generated field on the instability growth rate is not substantial at collisionless shocks in weakly magnetized media, including astrophysical shocks in GRBs. One should bear in mind, however, the long-term simulations of a shock show that the overall magnetic field strength keeps gradually increasing with time due to the cosmic rays driving the instability in the foreshock \citep{keshet+09}. Thus, the role of quasi-collisions may greatly increase if $\beta$ becomes small.

\section{Conclusions}

In this parer, we studied the role of pitch-angle scattering of particles in sub-Larmor-scale magnetic fields, referred to as `quasi-collisions', on the growth rate of the Weibel instability. The general formalism of such a non-linear effect has been presented. The results can describe the back-reaction of the self-generated magnetic fields on the instability growth rate in a deeply nonlinear regime, beyond the domain of applicability of the quasilinear theory. Hence, we colloquially refer to it as the `quasi-nonlinear'. The estimate of the magnitude of the effect for the foreshock conditions of collisionless shocks in weakly magnetized plasmas and astrophysical shocks in GRBs in particular is presented.

\acknowledgements

The author is grateful to the the Institute for Theory and Computation at Harvard University for support and hospitality and acknowledges DOE support via grant DE-SC0016368. 


\begin{thebibliography}{DUM}
%
\bibitem[Weibel(1959)]{weibel59} 
Weibel, E.S. 1959, \prl, 2, 83 
%
\bibitem[Fried(1959)]{fried59} 
Fried, B.~D.\ 1959, Physics of Fluids, 2, 337
%
\bibitem [Medvedev \& Loeb(1999)]{medvedev+99}
Medvedev, M. V., \& Loeb, A. \ 1999, \apj, 526, 697
%
\bibitem[Nishikawa et al.(2003)]{nishikawa+03} 
Nishikawa, K.-I., Hardee, P., Richardson, G., Preece, R., Sol, H., Fishman, G. J.\ 2003, \apj, 595, 555 
%
\bibitem[Silva et al.(2003)]{silva+03} 
Silva, L.~O., Fonseca, R.~A., Tonge, J.~W., Dawson, J. M., Mori, W. B., Medvedev, M. V.\ 2003, \apjl, 596, L121
%
\bibitem[Frederiksen et al.(2004)]{frederiksen+04}
Frederiksen, J.~T., Hededal, C.~B., Haugb{\o}lle, T., \& Nordlund, {\AA}.\ 2004, \apjl, 608, L13
%
\bibitem[Nishikawa et al.(2005)]{nishikawa+05} 
Nishikawa, K.-I., Hardee, P., Richardson, G.,  Preece, R., Sol, H., Fishman, G. J.\ 2005, \apj, 622, 927 
%
\bibitem[Spitkovsky(2008)]{spitkovsky+08} 
Spitkovsky, A.\ 2008, \apjl, 673, L39 
%
\bibitem[Nishikawa et al.(2009)]{nishikawa+09} 
Nishikawa, K.-I., Niemiec, J., Hardee, P.~E.,  Medvedev, M., Sol, H., Mizuno, Y., Zhang, B., Pohl, M., Oka, M., Hartmann, D. H.\ 2009, \apjl, 698, L10 
%
\bibitem[Keshet et al.(2009)]{keshet+09} 
Keshet, U., Katz, B., Spitkovsky, A., \& Waxman, E.\ 2009, \apjl, 693, L127 
%
\bibitem[Medvedev \& Zakutnyaya(2009)]{medvedev+z09}
Medvedev, M.~V., \& Zakutnyaya, O.~V.\ 2009, \apj, 696, 2269 
%
\bibitem[Kamble et al.(2014)]{kamble+14} 
Kamble, A., Soderberg, A.~M., Chomiuk, L., Margutti, R., Medvedev, M., Milisavljevic, D., Chakraborti, S., Chevalier, R., Chugai, N., Dittmann, J., Drout, M., Fransson, C., Nakar, E., Sanders, N. \ 2014, \apj, 797, 2 
%
\bibitem[Swisdak et al.(2008)]{swisdak+08}
Swisdak, M., Liu, Y.-H., \& Drake, J.~F.\ 2008, \apj, 680, 999
%
\bibitem[Liu \& Swisdak(2009)]{liu+09}
Liu, Y.-H., Swisdak, M., \& Drake, J.~F.\ 2009, Physics of Plasmas, 16, 042101
%
\bibitem[Tatarakis et al.(2003)]{tatarakis+03} 
Tatarakis, M., Beg, F.~N., Clark, E.~L., Dangor, A.~E., Edwards, R.~D., Evans, R.~G., Goldsack, T.~J., Ledingham, K.~W., Norreys, P.~A., Sinclair, M.~A., Wei, M.-S., Zepf, M., Krushelnick, K. \ 2003, \prl, 90, 175001 
%
\bibitem[Ren et al.(2004)]{ren+04}	
Ren, C., Tzoufras, M., Tsung, F.~S., Mori, W.~B., Amorini, S., Fonseca, R.~A., Silva, L.~O., Adam, J.~C., Heron, A. \ 2004, \prl, 93, 185004 
%
\bibitem[Huntington(2008)]{huntington+12} 
Huntington, C. M.,\ 2012, Ph.D.~Thesis.
%
\bibitem[Mondal et al.(2012)]{mondal+12}
Mondal, S., Narayanan, V., Ding, W.~J.,  Lad, A. D., Hao, B., Ahmad, S., Wang, W. M., Sheng, Z. M., Sengupta, S., Kaw, P., Das, A,, Kumar, G. R.\ 2012, Proceedings of the National Academy of Science USA, 109, 8011
%
\bibitem[Kugland et al.(2012)]{kugland+12} 
Kugland, N.~L., Ryutov, D.~D., Chang, P.-Y., Drake, R. P., Fiksel, G., Froula, D. H., Glenzer, S. H., Gregori, G., Grosskopf, M., Koenig, M., Kuramitsu, Y., Kuranz, C., Levy, M. C., Liang, E., Meinecke, J., Miniati, F., Morita, T., Pelka, A., Plechaty, C., Presura, R., Ravasio, A., Remington, B. A., Reville, B., Ross, J. S., Sakawa, Y., Spitkovsky, A., Takabe, H., Park, H.-S.\ 2012, Nature Physics, 8, 809 
%
\bibitem[Huntington et al.(2015)]{huntington+15} 
Huntington, C.~M., Fiuza, F., Ross, J.~S., Zylstra, A.~B., Drake, R.~P., Froula, D.~H., Gregori, G., Kugland, N.~L., Kuranz, C.~C., Levy, M.~C., Li, C.~K., Meinecke, J., Morita, T., Petrasso, R., Plechaty, C., Remington, B.~A., Ryutov, D.~D., Sakawa, Y., Spitkovsky, A., Takabe, H., Park, H.-S. \ 2015, \nat, 11, 2, 173
%
\bibitem[Park et al.(2015)]{park+15} 
Park, H.-S., Huntington, C.~M., Fiuza, F., Drake, R.~P., Froula, D.~H., Gregori, G., Koenig, M., Kugland, N.~L., Kuranz, C.~C., Lamb, D.~Q., Levy, M.~C., Li, C.~K., Meinecke, J., Morita, T., Petrasso, R., Pollock, B.~B., Remington, B.~A., Rinderknecht, H.~G., Rosenberg, M., Ross, J.~S., Ryutov, D.~D., Sakawa, Y., Spitkovsky, A., Takabe, H., Turnbull, D.~P., Tzeferacos, P., Weber, S.~V., Zylstra, A.~B. \ 2015, Physics of Plasmas, 22, 056311
%
\bibitem[Medvedev(2007)]{medvedev07} 
Medvedev, M.~V.\ 2007, \apss, 307, 245 
%
\bibitem[Medvedev(2008)]{medvedev08} 
Medvedev, M.\ 2008, Bulletin of the American Astronomical Society, 40, 3.42 
%
\bibitem[Medvedev \& Spitkovsky(2009)]{medvedev+s09} 
Medvedev, M.~V., \& Spitkovsky, A.\ 2009, \apj, 700, 956 
%
\bibitem[Medvedev et al.(2005)]{medvedev+05} 
Medvedev, M.~V., Fiore, M., Fonseca, R.~A., Silva, L.~O., \& Mori, W.~B.\ 2005, \apjl, 618, L75 
%
\bibitem[Bret et al.(2005)]{bret+05} Bret, A., Firpo, M.-C., \& Deutsch, C.\ 2005, \prl, 94, 115002 
%
\bibitem[Bret et al.(2008)]{bret+08} 
Bret, A., Gremillet, L., B{\'e}nisti, D., \& Lefebvre, E.\ 2008, \prl, 100, 205008 
%
\bibitem[Achterberg \& Wiersma(2007)]{achterberg+07a} Achterberg, A., \& Wiersma, J.\ 2007, \aap, 475, 1 
%
\bibitem[Achterberg et al.(2007)]{achterberg+07b} Achterberg, A., Wiersma, J., \& Norman, C.~A.\ 2007, \aap, 475, 19 
%
\bibitem[Lemoine \& Pelletier(2009)]{lemoine+09}
Lemoine, M., \& Pelletier, G.\ 2010, \mnras, 402, 321 
%
\bibitem[Medvedev et al.(2011)]{medvedev11} 
Medvedev, M.V., Frederiksen, J.T., Haugb\o lle, T., Nordlund, \AA. 2011, \apj, 737, 55
%
\bibitem[Fiuza et al.(2012)]{fiuza+12} 
Fiuza, F., Fonseca, R.~A., Tonge, J., Mori, W.~B., \& Silva, L.~O.\ 2012, \prl, 108, 235004 
%
\bibitem[Keenan \& Medvedev(2013)]{keenan+13} 
Keenan, B.~D., \& Medvedev, M.~V.\ 2013, \pre, 88, 013103 
%
\bibitem[Keenan et al.(2015)]{keenan+15} 
Keenan, B.~D., Ford, A.~L., \& Medvedev, M.~V.\ 2015, \pre, 92, 033104 
%
\bibitem[Keenan \& Medvedev(2016)]{keenan+16} 
Keenan, B.~D., \& Medvedev, M.~V.\ 2016, Journal of Plasma Physics, 82, 905820207 
%
\bibitem[Biswas \& Eswaran(2002)]{biswas+02}
Biswas, G. \& Eswaran, V. 2002, {\it Turbulent Flows: Fundamentals, Experiments and Modeling}, {IIT Kanpur series of advanced texts}. (CRC Press)
%
\bibitem[Honda(2004)]{honda04}
Honda, M.\ 2004, \pre, 69, 016401 
%
\bibitem[Califano, et al.(1997)]{califano+97}
Califano, F., Perogaro, F., \& Bulanov, S.V.\ 1997, \pre, 56, 963 
%
\end{thebibliography}

{

}

\end{document}